\documentclass[aps,prd,preprint,superscriptaddress,nofootinbib]{revtex4-1}
\usepackage{chay,graphicx,hyperref,color}

\newcommand{\ms}{\Bigl(\frac{\mu^2 e^{\gamma_{\mathrm{E}}}}{4\pi}\Bigr)^{\epsilon}}

\mathchardef\mhyphen="2D
\newcommand{\fms}[1]{{#1}\!\!\!/}

\newcommand{\mc}{\mathcal}
\newcommand{\mr}{\mathrm}
\newcommand{\mP}{\mathcal{P}}

\newcommand{\be}{\begin{equation}} 
\newcommand{\ee}{\end{equation}} 
\newcommand{\bea}{\begin{eqnarray}} 
\newcommand{\eea}{\end{eqnarray}} 

\newcommand{\ov}{\overline}

\newcommand{\n}{\overline{n}}
\newcommand{\nn}{\frac{\fms{\overline{n}}}{2}}

\newcommand{\blp}[1]{{\bf{#1}}_{\perp}}
\newcommand{\blpu}[1]{{\bf{#1}}^{\perp}}

\newcommand{\nnb}{\nonumber} 

\newcommand{\as}{\alpha_s} 

\newcommand{\eps}{\epsilon} 
 
\newcommand{\w}{\omega}

\begin{document}
\title{Threshold Factorization Redux}

\def\KU{Department of Physics, Korea University, Seoul 02841, Korea} 
\def\ibs{Fields, Gravity \& Strings, CTPU, Institute for Basic Science, Seoul 08826, Korea}
\def\Seoultech{Institute of Convergence Fundamental Studies and School of Liberal Arts, 
Seoul National University of Science and 
Technology, Seoul 01811, Korea}
\author{Junegone Chay}
\email[E-mail: ]{chay@korea.ac.kr}
\affiliation{\KU}
\affiliation{\ibs}
\author{Chul Kim}
\email[E-mail: ]{chul@seoultech.ac.kr}
\affiliation{\Seoultech}

\begin{abstract} \vspace{0.1cm}\baselineskip 3.0 ex 
We reanalyze the factorization theorems for Drell-Yan process and for deep inelastic scattering near threshold, as
constructed in the framework of the soft-collinear effective theory (SCET), from a new, consistent perspective.
In order to formulate the factorization near threshold in SCET, we should include an additional degree of freedom with small energy, collinear to the beam direction.
The corresponding collinear-soft mode is included to describe the parton distribution function (PDF) 
near threshold. The soft function is modified by subtracting the contribution of the collinear-soft modes 
in order to avoid double counting on the overlap region.  As a result, the proper soft function becomes infrared finite, and all 
the factorized parts are free of rapidity divergence. Furthermore, the separation of the relevant scales in each factorized part becomes manifest. 
We apply the same idea to the dihadron production in $e^+ e^-$ annihilation near threshold, and show that the resultant soft function 
is also free of infrared and rapidity divergences. 
\end{abstract}

\maketitle

\baselineskip 3.0 ex 

\section{Introduction}

Factorization theorems in which high-energy processes are divided into  hard, collinear, and soft parts are 
essential in providing precise theoretical predictions. 
Though it is difficult to probe the threshold region experimentally, it is theoretically both interesting and tantalizing to exploit the factorization near threshold
in Drell-Yan (DY) process and in deep inelastic scattering (DIS). 
A prominent distinction near threshold is that there exists nonvanishing soft interaction
since the real contribution does not cancel the virtual contribution completely due to the kinematic constraint near 
threshold~\cite{Sterman:1986aj,Catani:1989ne}. This distinctive feature near threshold was also discussed in 
Refs.~\cite{Manohar:2003vb, Chay:2005rz, Becher:2006mr,Becher:2007ty} in the framework of the soft-collinear effective theory 
(SCET)~\cite{Bauer:2000ew,Bauer:2000yr,Bauer:2001yt,Bauer:2002nz}.

%With this in mind, we can naively guess that the structure functions $F_{\mathrm{DY}}$ for DY process and 
%$F_1$ for DIS in the Breit frame, can be schematically written in a factorized form as
It is well established in full quantum chromodynamics (QCD) that the structure functions $F_{\mathrm{DY}}$ for DY process and 
$F_1$ for DIS can be schematically written in a factorized form as~\cite{Sterman:1986aj,Catani:1989ne}
\begin{eqnarray} \label{scheme}
F_{\mathrm{DY}}(\tau) &=&  H_{\mathrm{DY}}(Q,\mu) \cdot S_{\mathrm{DY}} (Q(1-z),\mu) \otimes f_{q/N_1} (x_1,\mu) \otimes f_{\bar{q}/N_2} (x_2,\mu),  \\
\label{sDIS}
F_1(x) &=&H_{\mathrm{DIS}}  (Q,\mu) \cdot J(Q\sqrt{1-x},\mu) \otimes S_{\mathrm{DIS}} (Q(1-x),\mu) \otimes f_{q/N} (x,\mu),
\end{eqnarray}
where the kinematical variables $\tau=q^2/s=Q^2/s$, $z=\tau/(x_1x_2)$, and $x=-q^2/2P\cdot q= Q^2/2P\cdot q$ are all close to 1.
Here $q^{\mu}$ is the hard momentum carried by a photon, $s$ is  the momentum squared of incoming hadrons in DY, and $P^{\mu}$ is 
the momentum of an incoming hadron.  The hard functions $H_{\mr{DY}}(Q)$ and $H_{\mathrm{DIS}}(Q)$ describe the physics at 
the large scale $Q$. 
The PDF $f_{i/N} (x,\mu)$ is the collinear part from the incoming hadrons, which represents the probability that a specific parton of type $i$ 
in a hadron $N$ has a longitudinal momentum fraction $x$.  The jet function $J(Q\sqrt{1-x},\mu)$ describes the energetic collinear particles 
in the final state in DIS.  Finally the soft functions $S_\mr{DY}(Q(1-z),\mu)$ and $S_{\mathrm{DIS}} (Q(1-x),\mu)$ describe the 
emission of soft particles. (Actually $S_{\mathrm{DIS}} (Q(1-x),\mu)=1$ to all orders in $\alpha_s$, as will be discussed later.)  
And `$\otimes$' implies an appropriate convolution.  

If we consider the factorization in SCET, the surmised form of the factorized structure functions in full QCD near threshold in
 Eqs.~(\ref{scheme}) and (\ref{sDIS}) looks reasonable at first glance, but there are delicate and discomfiting aspects.
 In fulll QCD, the PDF has the collinear divergences yielding the DGLAP equation through special definitions of the PDF or with other  techniques \cite{Sterman:1986aj}.  
  And the soft part with the eikonal cross sections is IR finite. In SCET, if we naively separate the collinear, soft modes, there 
appear IR and rapidity divergences in each factorized part. In this paper, we address this problem by considering additional modes required in
SCET near threshold. 

The issues on the factorization near threshold in SCET are summarized as follows:
First, since the incoming active partons take almost all the hadron momenta, 
the emission of additional collinear particles is prohibited. 
It means that only the virtual correction contributes to the collinear part~\cite{Fleming:2012kb,Fleming:2016nhs}.
Therefore, if we consider the collinear interaction alone, 
the PDF near threshold should be definitely different from the PDF away from threshold, where the latter includes the effect of real
gluon emissions.

Second, even though we accept Eqs.~(\ref{scheme}) and (\ref{sDIS}) and compute the factorized parts in SCET perturbatively, we 
encounter infrared (IR) divergences not only in the PDFs but also in the soft functions. The IR divergence in the PDF
can be safely absorbed in the nonperturbative part, but the IR divergence in the soft function is a serious problem since it destroys 
the factorization and prevents a legitimate resummation of large logarithms of $1-x$ or $1-z$.  

The existence of the IR divergence has  
been casually disregarded in the belief that the final physical result should be free of it.
But it was pointed out in Ref.~\cite{Chay:2012jr,Chay:2013zya} that the soft functions indeed
contain IR divergence by carefully separating the IR and ultraviolet (UV) divergences.  In Ref.~\cite{Chay:2012jr,Chay:2013zya}, 
we have suggested how some of the divergences can be transferred from the soft part to the collinear part to make the soft function IR finite, 
while the collinear part reproduces the PDF  near threshold.  Basically this amounts to
reshuffling divergences based on physics near threshold, but it was difficult to explain how the scale dependence in each part can be established 
consistently to resum large logarithms. For example, the typical scale for the PDF is $\mu \sim \Lambda_{\mathrm{QCD}}$ or larger, and 
the relevant scale to the soft function is $\mu \sim Q (1-z)$ or $Q (1-x)$. 

Third, as we will see later, the soft parts in Eqs.~(\ref{scheme}) and (\ref{sDIS}) include the rapidity divergence. 
The rapidity divergence  arises when the product of the lightcone momenta remains constant, while 
each component goes to zero or infinity~\cite{Chiu:2011qc,Chiu:2012ir}. 
Though the rapidity divergence exists in each factorized part, the scattering cross section, which is a convolution of the factorized parts, is free 
of the rapidity divergence. 
However, the cancellation of the rapidity divergence occurs only when the invariant masses of the different modes are of the same order. Near threshold, the invariant masses of the collinear particles and the soft particles are different and there is no reason for the rapidity divergence to cancel in the sum of the collinear and the soft parts.

%However, the anomalous dimensions of the factorized parts are IR divergent, which is hard to assimilate considering
%the idea of the renormalization group equation.

The naive extension of the factorized form in Eqs.~(\ref{scheme}) and (\ref{sDIS}) to SCET contains all these problems. 
And the questions are how we can obtain a consistent factorization formula near threshold, and how it can be connected to the factorization 
away from threshold. The predicament can be resolved by noticing that, near threshold, there is an additional degree of freedom with 
small energy, collinear to the beam direction. The small energy scale is given by $\omega=Q(1-z)$ in DY process or 
$\omega =Q(1-x)$ in DIS with $z\sim x \sim 1$, and lies  between the large scale $Q$ and the 
low scale $\Lambda_{\mathrm{QCD}}$. The main points of our paper are to identify the new degrees of freedom, to incorporate them in the 
definition of the PDF and the soft functions, to calculate the perturbative corrections at order $\alpha_s$, and to show that we obtain 
the proper factorization near threshold. 

The high-energy processes including the threshold region can be efficiently described by SCET.  
%\cite{Bauer:2000ew,Bauer:2000yr,Bauer:2001yt}.  
In SCET, the $n$-collinear momentum 
scales as $(\overline{n}\cdot p, n\cdot p, p_{\perp}^{\mu}) \sim Q (1,\lambda^2, \lambda)$, where $Q$ is the large energy scale and $\lambda$
is the small parameter for power counting in SCET. The $\overline{n}$-collinear momentum scales as $Q(\lambda^2,1,\lambda)$.
%while the soft momentum scales as $Q(\lambda,\lambda,\lambda)$. 
Here $n^{\mu}$ and $\overline{n}^{\mu}$ denote the lightcone vectors satisfying
$n^2=0$, $\overline{n}^2 =0$ and $n\cdot \overline{n}=2$.
In order to describe the threshold region, from the $n$- and $\n$-collinear interactions
we decouple the $n$-collinear-soft (csoft) and the $\overline{n}$-csoft modes respectively, which scale as
\begin{equation} 
p^{\mu}_{n,cs} \sim \omega(1,\alpha^2,\alpha),  \
p^{\mu}_{\bar{n},cs}  \sim  \omega (\alpha^2,1,\alpha).
\end{equation}
Here the power-counting parameter $\alpha$ for the csoft modes satisfies the relation $\omega \alpha \sim Q\lambda$, such that the collinear and 
csoft particles have $p^2 \sim \omega^2 \alpha^2 \sim Q^2 \lambda^2$. The additional partition of the collinear modes depends 
on the new scale introduced 
near threshold.  The csoft modes are soft since the overall scale is governed by the small scale $\omega$, but the momentum components 
scale like collinear momenta. The nomenclature for the collinear-soft modes varies, as in
the  collinear-soft (csoft) modes \cite{Bauer:2011uc,Procura:2014cba}, the coft modes \cite{Becher:2015hka}, and the soft-collinear modes \cite{Chien:2015cka}, referring
to the modes with similar momentum scaling in  different situations. Here we will simply call these modes the csoft modes.

The important feature near threshold is that the incoming 
active parton cannot emit real collinear particles, but the particles in the csoft modes can be emitted. 
And we define the PDF near threshold including the csoft modes. The new definition of the PDF covers the threshold region
as well as the region away from threshold since the effect of the csoft modes away from threshold is cancelled to all orders, while it correctly describes
the PDF near threshold. To avoid double counting on the overlap region, the contribution of the csoft modes should be subtracted
from the soft part to obtain the soft function. Here the soft modes near threshold scale as $p_s \sim \w(1,1,1)$, hence the csoft mode can be also considered to be a subset of the soft mode.  

The effect of the csoft modes in the collinear and soft modes is more interesting 
when we  consider the details of the higher order calculations. Without the csoft modes,
both the soft part and the PDF contain the IR and rapidity divergences.  But when the csoft modes
are subtracted from the soft part, the resultant soft functions are free of the IR and rapidity divergences, and the PDF is free of rapidity divergence 
when the contribution of the csoft modes is included.  

The structure of the paper is as follows: In Sec.~\ref{csofti}, the main idea of incorporating the csoft modes is presented in SCET. The PDF
and the soft functions are defined near threshold in DY and in DIS processes. In Sec.~\ref{scom}, the soft functions and the PDF are 
computed at order $\alpha_s$ with the csoft modes.   In Sec.~\ref{dihadron}, we consider the factorization of 
the dihadron production in $e^+ e^-$ annihilation near threshold, in which the effect of the csoft modes is included in the fragmentation functions.
Finally we conclude in Sec.~\ref{conc}.

\section{Factorization near threshold with the appropriate PDF\label{csofti}}

\subsection{Extension of the PDF to threshold}
\label{PDFsec}

The main issue in constructing the PDF near threshold is how to implement the tight kinematic constraint, and how to relate it to the PDF  away from threshold. Near threshold, the incoming partons cannot emit real collinear particles, hence only the virtual corrections contribute.
On the other hand, the emission of the csoft modes is allowed. Therefore we start from defining the PDF near threshold by subdividing the collinear field into the collinear and the csoft modes.
In Ref.~\cite{Dai:2017dpc}, the decomposition of the collinear and the csoft modes has been performed in order to describe a fragmenting process to a jet with a large momentum fraction $z$. It can be adopted in defining the PDF near threshold.

We first decouple the soft mode $\sim Q\zeta(1, 1, 1)$  near threshold from the collinear mode, 
where $\zeta$ is a small parameter to characterize $1-z$ or $1-x$. 
This is obtained by redefining the 
collinear fields in terms of the soft Wilson line \cite{Bauer:2001yt} as
\begin{equation} \label{yn}
\xi_n \rightarrow Y_n  \xi_n, \ A_n^{\mu} \rightarrow Y_n  A_n^{\mu} Y_n^{\dagger},
\end{equation}
where the soft Wilson line $Y_n$ is given by
\begin{equation} 
\label{Yn}
Y_n  (x) =P\exp\Bigl[ig \int_x^{\infty} ds n\cdot A_s (sn)\Bigr].
\end{equation}

Then we extract
the csoft mode $\sim Q\zeta (1,\alpha^2,\alpha)$ in the collinear sector. That is,  the collinear gluon $A_n^{\mu}$ is further decomposed into
$A_n^{\mu} \rightarrow A_n^{\mu} +A_{n,cs}^{\mu}$. The resultant collinear mode scales with the large energy $Q$, while the csoft mode scales
with $Q\zeta$.
After the decomposition, the covariant derivative can be written as  
$iD^{\mu} = iD_c^{\mu} +iD_{cs}^{\mu} =\mathcal{P}^{\mu} +gA_n^{\mu} +i\partial^{\mu} +g A_{cs}^{\mu}$. Here $\mathcal{P}^{\mu}$ 
($i\partial^{\mu}$) is the operator extracting the collinear (csoft) momentum, which applies only to the collinear (csoft) operator.

The collinear quark distribution function, which is the PDF away from threshold, is defined as
\begin{eqnarray} 
f_{q/N} (x,\mu) &=& \langle N (P_+) | \overline{\chi}_n \frac{\overline{\FMslash{n}}}{2} \delta (x  P_+ -\mathcal{P}_+)\chi_n|N(P_+) \rangle \nnb \\
\label{cdfn}
 &=& \langle N (P_+) | \overline{\xi}_n \frac{\overline{\FMslash{n}}}{2} \delta (x  P_+  -\n\cdot iD_c)\xi_n|N(P_+) \rangle,
\end{eqnarray}
where $\mathcal{P}_+ \equiv \n\cdot \mP$ is the operator extracting the largest momentum component from the collinear field. The average
over spin and color is included in the matrix element. The combination
$\chi_n = W_n^{\dagger} \xi_n$ is the collinear gauge-invariant block with the collinear Wilson line $W_n$ and the collinear quark field $\xi_n$. 

 For the proper treatment of the PDF near threshold, we need to include the csoft mode, which describes the emission along the beam direction. It is
implemented by replacing $\n\cdot iD_c$ with $\n\cdot iD = \n\cdot iD_c + \n\cdot iD_{cs}$. Then 
we define the PDF as
\begin{equation} \label{pdfgen}
\phi_{q/N} (x,\mu) =\langle N  | \overline{\xi}_n \frac{\overline{\FMslash{n}}}{2} \delta (x  P_+  -
\overline{n}\cdot iD)\xi_n|N \rangle,
\end{equation}
which covers all the regions, near and away from threshold.
Note that the expression for $\phi_{q/N}$  is invariant under the collinear and csoft gauge transformations.  
In order to show the gauge invariance order by order in power counting in a manifest way \cite{Bauer:2003mga}, 
we redefine the collinear gluon as
\begin{equation}
A_n^{\mu} =\hat{A}_n^{\mu} + \hat{W}_n [iD_{cs}^{\mu}, \hat{W}_n^{\dagger}],
\end{equation}
where the collinear Wilson line $\hat{W}_n$ is expressed in terms of the newly defined collinear gluon field $\hat{A}^{\mu}$. 
Then the covariant derivative can be written as
\begin{equation}
iD^{\mu} = iD_c^{\mu} +W_n iD_{cs}^{\mu} W_n^{\dagger},
\end{equation}
where the hats in $W_n$ and $A_n$ are removed for simplicity. Then the delta function in Eq.~(\ref{pdfgen}) is written as
\begin{equation}
\delta (x  P_+ - \overline{n}\cdot iD) =W_n \delta (x  P_+  -\mathcal{P}_+
- \overline{n}\cdot iD_{cs})W_n^{\dagger}.
\end{equation}

We can decouple the csoft interaction from the collinear part by redefining the collinear field as
\begin{equation} \label{csdec}
\xi_n \rightarrow Y_{n,cs} \xi_n, \ A_n^{\mu} \rightarrow Y_{n,cs} A_n^{\mu} Y_{n,cs}^{\dagger},
\end{equation}
which is similar to the decoupling of the soft interaction in Eq.~(\ref{yn}).
Here the csoft Wilson line $Y_{n,cs}$ is defined as
\begin{equation}
Y_{n,cs} (x) =P\exp\Bigl[ig \int_x^{\infty} ds n\cdot A_{cs} (sn)\Bigr],
\end{equation}
and $Y_{\bar{n},cs}$ is obtained by switching $n\leftrightarrow \overline{n}$.
Using the relation $\overline{n}\cdot iD_{cs} = Y_{\bar{n},cs} \overline{n}\cdot i\partial Y_{\bar{n},cs}^{\dagger}$, the final expression for the PDF is given by
\be 
\label{pdfth}
\phi_{q/N} (x,\mu) = \langle N | \overline{\chi}_n \frac{\overline{\FMslash{n}}}{2} Y_{n,cs}^{\dagger}Y_{\bar{n},cs} 
\delta (x  P_+ - \mP_+ -i\partial_+)Y_{\bar{n},cs}^{\dagger} Y_{n,cs}\chi_n|N\rangle_. 
\ee

Note that this new definition of the PDF is also valid in the region away from threshold. 
In this case the term $i\partial_+$ in the delta function is much smaller than $xP_+ - \mP_+$, and it can be safely neglected. 
Then, due to the unitarity of the csoft Wilson line, the effect of the csoft modes cancels to all orders. Hence we can recover Eq.~(\ref{cdfn}) and 
describe the PDF away from threshold with the collinear interactions only. 
Near threshold, we can put the label momentum in Eq.~(\ref{pdfth}) as $\mP_+ = P_+$, and obtain the PDF as 
\be
\label{pdflx} 
\phi_{q/N} (x\to 1,\mu) = \langle N | \overline{\chi}_n \frac{\overline{\FMslash{n}}}{2} Y_{n,cs}^{\dagger}Y_{\bar{n},cs} 
\delta ((1-x)P_+  + i\partial_+)Y_{\bar{n},cs}^{\dagger} Y_{n,cs}\chi_n|N\rangle_.
\ee 
The same result near threshold has been also derived in $\mr{SCET_+}$ approach~\cite{Hoang:2015iva}.
Therefore the expression in Eq.~(\ref{pdfth}) can be regarded as the definition of the PDF over all kinematic regions.
Using the expression in Eq.~(\ref{pdflx}), we can calculate the PDF near threshold at the parton level,  i.e., $\phi_{q/q}$.
As we will see later, the calculation exactly reproduces the standard PDF  in the limit $x\to 1$ and it satisfies the 
Dokshitzer-Gribov-Lipatov-Altarelli-Parisi (DGLAP) evolution. 

From Eq.~(\ref{pdflx}), the fluctuation of the csoft mode is estimated as $p_{cs}^2\sim \Lambda_{\mr{QCD}}^2$. 
Therefore this mode scales as $p_{cs}^{\mu} \sim Q\zeta(1,\alpha^2,\alpha)$ with $\alpha = \Lambda_{\mr{QCD}}/(Q\zeta)$. 
When we discuss the decomposition of the collinear and the csoft modes below Eq.~(\ref{Yn}), the csoft mode can be regarded as a subset 
of the collinear mode.  Hence the collinear mode would scale as $Q(1,\lambda^{\prime2},\lambda^{\prime})$ 
with $\lambda^{\prime} \sim \zeta^{1/2} \alpha$. 
And the offshellness is given by $p_c^{ 2} \sim \Lambda _{\mr{QCD}}^2/\zeta$, which is much larger than the typical hadronic scale squared. 
However, near threshold this collinear mode contributes to the PDF only through the virtual corrections 
without any specific scale. So there is no impact on integrating out the mode $\sim Q(1,\lambda^{\prime2},\lambda^{\prime})$ 
and we can scale it down to $p_c \sim Q(1,\lambda^2,\lambda)$ with $\lambda \sim \Lambda _{\mr{QCD}}/Q$.  Then the collinear mode 
at the lower scale has the offshellness $p_c^2 \sim \Lambda _{\mr{QCD}}^2$.

\subsection{Prescription of the factorization near threshold}

With the new definition of the PDF, the naive factorization formulae which are schematically given in Eqs.~(\ref{scheme}) and (\ref{sDIS}) 
can be cast into an appropriate form near threshold. In order to construct correct factorization theorems near threshold, we take the following steps: 
\begin{itemize}
\item After integrating out the hard interactions, we construct the naive factorization formalism by decomposing the collinear and soft interactions.
\item Next we decouple the csoft mode from the collinear mode, and express the PDF in terms of the collinear and the csoft fields.\footnote{\baselineskip 3.0ex  The csoft modes are included in the definition of the PDF in full QCD, but in SCET, we have to devise these modes explicitly and add them to the PDF.   }
\item We define the soft functions by subtracting the csoft contributions to avoid double counting.  
\end{itemize}

The derivation of the naive factorization for the structure functions in SCET is not repeated here. 
Instead we refer to Ref.~\cite{Chay:2013zya}, where the details of the derivation are presented and the naively factorized results 
are shown in Eqs.~(2.13) and (2.33) for DY and DIS respectively. 

The naive factorization for the structure function in DY process is written as 
\begin{eqnarray} 
F_{\mathrm{DY}} (\tau) &=& -N_c \int \frac{d^4 q}{(2\pi)^4} \theta(q_0) \delta (Q^2 - s\tau) \int d^4 x e^{-iq\cdot x} 
\langle N_1 N_2|J_{\mu}^{\dagger} (x) J^{\mu} (0)|N_1 N_2\rangle \nonumber \\
\label{fdyn}
&=& H_{\mathrm{DY}}(Q,\mu) \int^1_{\tau} \frac{dz}{z} \tilde{S}_{DY} (Q(1-z),\mu) \int^1_{\tau/z} f_{q/N_1} (y) 
f_{\bar{q}/N_2} \bigl(\frac{\tau}{zy}\bigr),
\end{eqnarray}
where $N_c$ is the number of colors, $J_{\mu}$ is the electromagnetic current, and $Q^2 =q^2$ is the invariant mass squared of the lepton pair. 
$\tau=Q^2/s$ and  $z= Q^2/\hat{s}$, where $\hat{s}$ is the center-of-mass energy squared for the incoming partons. The cross section is given by 
$d\sigma/d\tau = \sigma_0 F_{\mathrm{DY}} (\tau)$ where $\sigma_0 = 4\pi \alpha^2 Q_f^2/(3N_c Q^2)$ is the Born cross section for 
the quark flavor $f$ with the electric charge $Q_f$.   The naive soft function $\tilde{S}_{DY}$ is defined as 
\begin{equation} \label{nsoftdy}
\tilde{S}_{\mathrm{DY}} (Q(1-z),\mu) = \frac{1}{N_c} \mathrm{Tr}  \langle 0|Y_n^{\dagger} Y_{\bar{n}} 
\delta \Bigl( 1-z+\frac{2i\partial_0}{Q} \Bigr)
Y_{\bar{n}}^{\dagger} Y_n|0\rangle.
\end{equation}
We call this naive soft function since the subtraction of the csoft modes is not included yet.

The naive factorization for the structure function $F_1$ in the Breit frame is given by
\begin{eqnarray}
F_1 (x) &=& 
-\frac{1}{4\pi} \sum_X (2\pi)^4 \delta^{(4)} (q+P-p_X)\langle N|J_{\mu}^{\dagger} |X\rangle \langle X|J^{\mu}|N\rangle 
\nonumber \\
\label{fdisn} 
&=& H_{\mathrm{DIS}} (Q,\mu) \int_x^1 \frac{dz}{z} f_{q/N} \Bigl(\frac{x}{z},\mu\Bigr) \int_z^1 \frac{dy}{y} J(Q\sqrt{1-y},\mu) 
\tilde{S}_{\mathrm{DIS}} \Bigl(Q(1-\frac{z}{y}),\mu \Bigr),
\end{eqnarray}
where $P^{\mu}=P_+ n^{\mu}/2$ is the momentum of the hadron $N$ along the beam and the final-state jet function in the $\overline{n}$ 
direction is defined as~\cite{Bauer:2001yt} 
\begin{equation}
\sum_{X_{\bar{n}}} \chi_{\bar{n}} |X_{\bar{n}}\rangle \langle X_{\bar{n}}|\overline{\chi}_{\bar{n}} 
=\frac{\FMslash{\overline{n}}}{2} \int \frac{d^4 p_{X_{\bar{n}}}}{(2\pi)^3} n\cdot P_{X_{\bar{n}}} J( p_{X_{\bar{n}}}). 
\end{equation}
Here  the naive soft function $\tilde{S}_{\mathrm{DIS}}$ is given as
\begin{equation} \label{dissoft}
\tilde{S}_{\mathrm{DIS}} (Q(1-z),\mu ) =\frac{1}{N_c} \mr{Tr} \langle 0|\mathrm{Tr} Y_n^{\dagger} Y_{\bar{n}} \delta \Bigl(1-z +\frac{\overline{n}\cdot  i\partial}{Q}\Bigr) Y_{\bar{n}}^{\dagger} Y_n |0\rangle.
\end{equation}

Based on the naive factorization in Eq.~(\ref{fdyn}) and (\ref{fdisn}), we construct the proper factorization theorem as follows: 
As shown in section.~\ref{PDFsec}, we decompose the collinear and the csoft modes in the collinear sector and express the PDF 
in Eq.~(\ref{pdflx}). This procedure can be achieved by replacing the collinear PDF $f_{i=q,\bar{q}/N}$ with the appropriate PDF $\phi_{i/N}$. 
Then we note that the csoft modes scaling as $p_{n,cs} \sim Q\zeta (1,\alpha^2,\alpha)$ and 
$p_{\bar{n},cs} \sim Q\zeta (\alpha^2,1,\alpha)$ are also the subsets of the soft mode $\sim Q\zeta(1,1,1)$, 
Therefore, in order  to avoid  double counting for the  overlap region, the csoft contribution should be subtracted from the naive soft function.
 
Finally the correct factorization theorem for DY and DIS processes are given as 
\bea 
\label{fdyc} 
F_{\mr{DY}} (\tau) &=& H_{\mathrm{DY}}(Q,\mu) \int^1_{\tau} \frac{dz}{z} S_{DY} (Q(1-z),\mu) \int^1_{\tau/z} \phi_{q/N_1} (y) \phi_{\bar{q}/N_2} \bigl(\frac{\tau}{zy}\bigr), \\
\label{fdisc}
F_{\mr{DIS}} (x) &=& H_{\mathrm{DIS}} (Q,\mu) \int_x^1 \frac{dz}{z} J(Q\sqrt{1-z},\mu)\phi_{q/N} \Bigl(\frac{x}{z},\mu\Bigr).
\eea
Note that there is no correlation between $\phi_{q/N_1}$ and $\phi_{\bar{q}/N_2}$ since the $n$-collinear(-csoft) and $\n$-collinear(-csoft) fields 
do not interact with each other at leading order in the power counting of the collinear(-csoft) limit.\footnote{\baselineskip 3.0ex
A counterexample might be Glauber interactions, that are not considered in this paper. We refer to Refs.~\cite{Rothstein:2016bsq,Schwartz:2017nmr}
for details. However  the Glauber gluon problem in full QCD was solved in Refs.~\cite{Collins:1985ue,Collins:1988ig} 
for the DY process, and it is  simpler in DIS.}
Here $S_{DY}$ is the soft function for DY after subtracting the csoft contribution from the naive soft function. 
The subtracting procedure follows the basic idea of the zero-bin subtraction~\cite{Manohar:2006nz}, which can be legitimately applied to the csoft sector~\cite{Lustermans:2016nvk}.
As a result the soft function 
contains neither IR nor rapidity divergence contrary to the naive soft function. 
In sec.~\ref{scom} we will see the details of the  computation for the soft function at order $\alpha_s$. 
When we compare Eq.~(\ref{fdisc}) with Eq.~(\ref{fdisn}), we see that $\tilde{S}_{\mr{DIS}}$ is not present in the final factorization theorem 
since the csoft contribution cancels  $\tilde{S}_{\mr{DIS}}$ to all orders in $\alpha_s$.

At tree level, both $S_{\mr{DY}}$ in Eq.~(\ref{fdyc}) and the naive soft function are normalized to $\delta(1-z)$.
At order $\alpha_s$, $S_{DY}$ is given as
\be
\label{nlosdys} 
S_{\mathrm{DY}}^{(1)}  = \tilde{S}_{\mathrm{DY}}^{(1)} - S_{cs}^{(1)} - S_{\overline{cs}}^{(1)},
\ee
where the superscript denotes the order in $\alpha_s$.
And $S_{cs}$ and $S_{\overline{cs}}$ are the $n$-csoft and $\n$-csoft contributions respectively, which are given as 
\bea
\label{Sncs} 
S_{cs} &=& \frac{1}{N_c} \mathrm{Tr}  \langle 0|Y_{n,cs}^{\dagger} Y_{\bar{n},cs} 
\delta \Bigl( 1-z+\frac{\n\cdot i\partial}{Q} \Bigr) Y_{\bar{n},cs}^{\dagger} Y_{n,cs}|0\rangle, \\
\label{Snbcs} 
S_{\ov{cs}} &=& \frac{1}{N_c} \mathrm{Tr}  \langle 0|Y_{n,\ov{cs}}^{\dagger} Y_{\bar{n},\ov{cs}} 
\delta \Bigl( 1-z+\frac{n\cdot i\partial}{Q} \Bigr) Y_{\bar{n},\ov{cs}}^{\dagger} Y_{n,\ov{cs}}|0\rangle.
\eea
Here $Y_{n,cs}$ and $Y_{n,\ov{cs}}$ are the csoft Wilson lines in terms of the $n$-csoft and $\n$-csoft gluons respectively. 
Note that $2i\partial_0$ in the argument of the naive soft function in Eq.~(\ref{nsoftdy}) is replaced by
$\n\cdot i\partial$ ($n\cdot i\partial$) in $S_{cs}$ ($S_{\ov{cs}} $) according to the power counting.

In order to specify the soft region completely in Eq.~(\ref{nlosdys}), we may introduce and add the contribution of the so-called `soft-soft (ssoft)' mode scaling as $p_{ss} \sim Q\zeta(\alpha^2,\alpha^2,\alpha^2)$. But the ssoft mode does not contribute 
to Eq.~(\ref{nlosdys}).
In general we can divide the full soft region into the `hard-soft (hsoft)', the csoft, and the ssoft regions. This division and the partition of the full soft 
region are similar to the procedure in constructing the hard, collinear, and soft modes in SCET from QCD. The only difference is that the large energy 
$Q$ in the full theory  is replaced with $Q\zeta$ here. Therefore,  we can systematically factorize the full soft modes into the hsoft, csoft and ssoft
degrees of freedom. In this respect, $S^{(1)}_{\mr{DY}}$ in  Eq.~(\ref{nlosdys}) can be regarded as the one-loop correction to the hsoft function obtained from the matching between the full soft and the csoft contributions. Since the csoft modes can reproduce the low energy behavior of the full soft function as the collinear modes do from the full theory, we argue that the hsoft function remains  IR finite at higher orders.

For the soft function in DIS, the nonzero csoft contribution only comes from $S_{cs}$ in Eq.~(\ref{Sncs}). 
And it is the same as the naive soft function, as well as the csoft contribution to the PDF in Eq.~(\ref{pdflx}). 
At order $\alpha_s$, $S_{\mr{DIS}}^{(1)} =  
\tilde{S}_{\mr{DIS}}^{(1)} - S_{cs}^{(1)} = 0$, and $S_{\mr{DIS}}(1-z)$ remains as $\delta(1-z)$. 
In fact, $\tilde{S}_{\mr{DIS}}$ and $S_{cs}$ are the same to all orders in $\alpha_s$, hence the soft function for DIS  
is given by $\delta (1-z)$ to all orders.

\section{The soft functions and the PDF near threshold \label{scom}}

We compute the soft functions and the PDF explicitly at one loop in order to verify the statements in the previous section.
We regulate the UV divergence using the dimensional regularization with $D=4-2\eps$ and the $\overline{\mr{MS}}$ scheme. 
We introduce the fictitious gluon mass $m_g$ to regulate the IR divergence. We also consider the rapidity 
divergence~\cite{Chiu:2011qc,Chiu:2012ir}, which appear as the loop momentum $k_+$ or $k_-$ goes to infinity while $k_+ k_-$ remains finite. 
We employ the Wilson lines as~\cite{Chiu:2012ir} 
\begin{equation}
W_n = \sum_{\mathrm{perms.}} \exp\Bigl[-\frac{g }{\overline{n}\cdot \mathcal{P}} 
\frac{|\overline{n}\cdot \mathcal{P}_g|^{-\eta}}{\nu^{-\eta}}
\overline{n}\cdot A_n\Bigr], \ Y_n = \sum_{\mathrm{perms.}} \exp \Bigl[-\frac{g}{n\cdot \mathcal{P}} 
\frac{|2\mathcal{P}_z|^{-\eta/2}}{\nu^{-\eta/2}} n\cdot A_s\Bigr],
\end{equation}
and the rapidity divergence appears as poles in $\eta$.
The Wilson lines in the $\overline{n}$ direction can be obtained by switching $n$ and $\overline{n}$. For the csoft mode, 
we use the form of the soft Wilson line, but the soft field is replaced by the csoft field.

\subsection{The soft functions near threshold}

We first consider the soft function for  DY process near threshold. 
The naive soft function is defined in Eq.~(\ref{nsoftdy}), and the correct soft function can be obtained through the csoft subtractions, 
given by Eq.~(\ref{nlosdys}). In order to see the scale dependence clearly, we introduce the dimensionful soft function, which is given as 
\be
\tilde{\mc{S}}_{DY}(\omega,\mu) = \frac{1}{Q}  S_{DY} (Q(1-z),\mu) =\frac{1}{N_c} \mathrm{Tr}  \langle 0|Y_n^{\dagger} Y_{\bar{n}} 
\delta \Bigl(\omega +2i\partial_0 \Bigr)
Y_{\bar{n}}^{\dagger} Y_n|0\rangle,
\ee
where $\omega = Q(1-z)$. 

The soft virtual contribution at one loop is given as 
\begin{eqnarray}
\label{MsV}
\mc{M}_{S}^V  &\equiv& \bar{\mc{M}}_{S}^V\cdot \delta(\w) = -4ig^2 C_F \ms \nu^{\eta}  \delta (\omega)\int \frac{d^Dk}{(2\pi)^D} \frac{|k_+ -k_-|^{-\eta}}{(k^2 -m_g^2) 
k_+ k_-}    \\
&=& \frac{\alpha_s C_F}{\pi}  \delta (\omega)\Biggl[\frac{1}{\eps^2} +\frac{1}{\eps} \Bigl( \ln \frac{\mu^2}{m_g^2} 
-\ln \frac{\nu^2}{m_g^2}\Bigr) -\frac{2}{\eta} \Bigl( \frac{1}{\eps} +\ln \frac{\mu^2}{m_g^2}  \Bigr)  
 +\frac{1}{2} 
\ln^2 \frac{\mu^2}{m_g^2} -\ln \frac{\mu^2}{m_g^2} \ln \frac{\nu^2}{m_g^2} -\frac{\pi^2}{12}\Biggr]_, \nonumber
\end{eqnarray}
where $k_+ =\overline{n}\cdot k$ and $k_-=n\cdot k$. 
The real gluon emission at order $\alpha_s$ is given as
\begin{eqnarray}
\mc{M}_{S,\mr{DY}}^R &=& \frac{\alpha_s C_F}{\pi} \frac{(\mu^2 e^{\gamma_E})^{\eps}}{\Gamma(1-\eps)} \nu^{\eta} \int \frac{dk_+dk_-}{k_+k_-}(k_+k_--m_g^2)^{-\eps}|k_+ - k_-|^{-\eta} \delta (\w -k_+ -k_- ) \Theta(k_+k_--m_g^2)\nonumber \\
&=& \frac{\alpha_s C_F}{\pi}  \int\frac{dk_+ dk_-}{k_+k_-} \delta (\w -k_+ -k_- ) \Theta(k_+k_--m_g^2)
\nonumber \\
\label{MRDY}
&=& \frac{\alpha_s C_F}{\pi}  \Biggl[\delta (\omega) \Bigl(\frac{1}{2} \ln^2 \frac{\Lambda^2}{m_g^2} -\frac{\pi^2}{6}\Bigr)
+\Bigl[\frac{2}{\omega} \ln \frac{\omega^2}{m_g^2}\Bigr]_{\Lambda} \Biggr]_.
\end{eqnarray}
Here $\Theta$ is the step function, and  
we put $\eps = \eta=0$ since the integral has neither the UV nor the rapidity divergence. The final result is expressed in terms of the 
$\Lambda$-distribution. It is defined as
\begin{equation}
\label{Ladist}
\int_0^L d\omega [g (\omega)]_{\Lambda} f(\omega) = \int_0^L d\omega g(\omega) f(\omega) - \int_0^{\Lambda} d\omega g(\omega) f(0),
\end{equation}
where $f$ is a smooth function at $\omega =0$. In defining the $\Lambda$-distribution, $\Lambda$ is an arbitrary scale larger than $\omega$,
but it suffices that $\Lambda$ is slightly larger than $\omega$.

\begin{figure}[t] 
\begin{center}
\includegraphics[height=8cm]{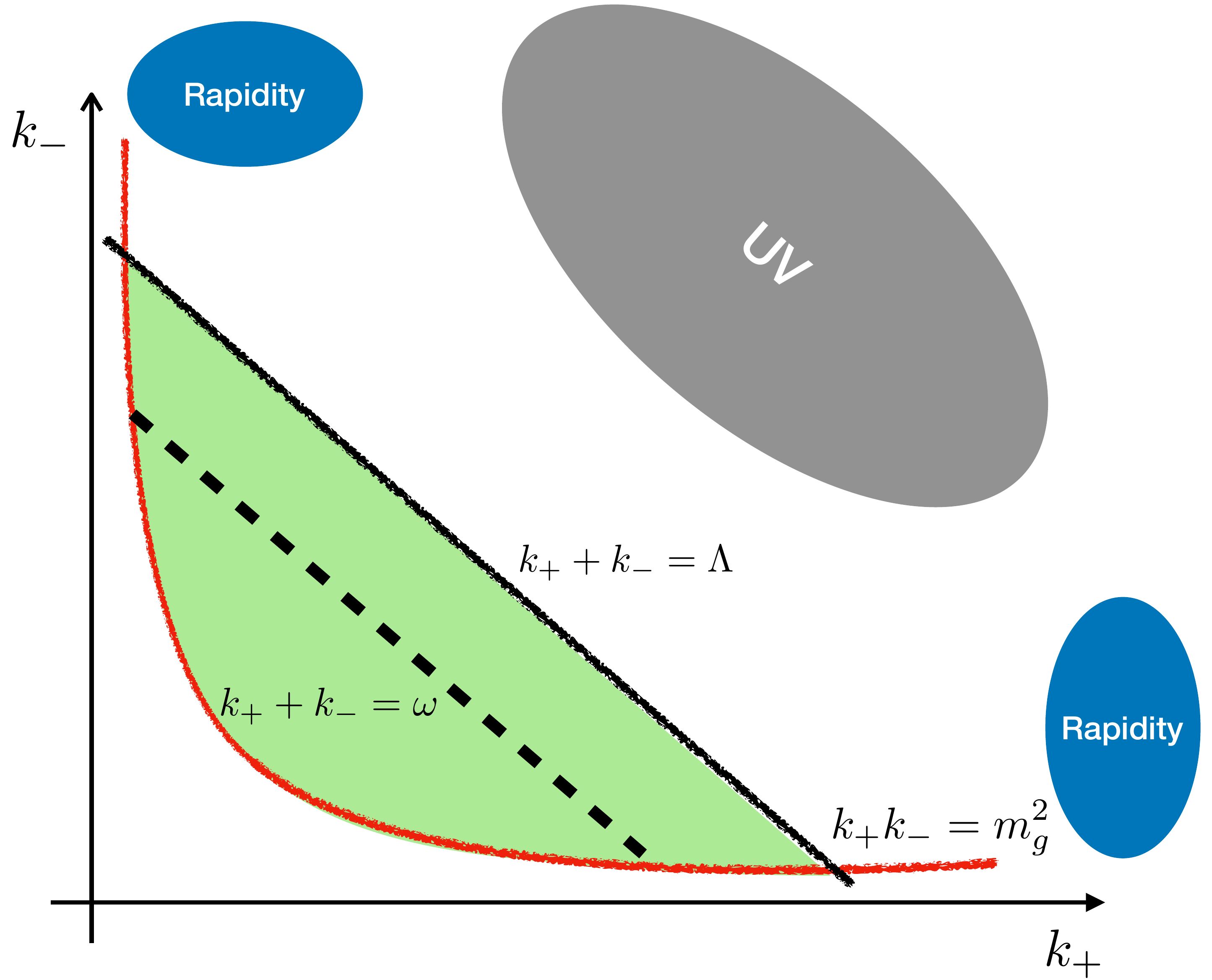}
\end{center}  
\vspace{-0.5cm}
\caption{\baselineskip 3.0 ex 
The phase space for the real gluon emission in the naive DY soft function after integrating over $\blp{k}^2$. The constraint specified by the
delta function is shown as the dashed line $k_++k_-=\w$.  The green region denotes the region of integration for the part proportional 
to $\delta(\w)$ with the $\Lambda$-distribution. The regions where the UV and the rapidity divergences arise are schematically illustrated. 
\label{phaseDY}}
\end{figure}

In obtaining the final result in Eq.~(\ref{MRDY}) with the $\Lambda$-distribution, we write $\mc{M}_{S,\mr{DY}}^R$ as 
\begin{equation} 
\label{epLdist}
\mc{M}_{S,\mr{DY}}^R (\omega) = \delta (\omega) \int_0^{\Lambda} d\omega^{\prime} \mc{M}_{S,\mr{DY}}^R (\omega') 
+[\mc{M}_{S,\mr{DY}}^R (\omega \neq 0)]_{\Lambda}.
\end{equation}
Note that this expression is independent of $\Lambda$. But the scale $\Lambda$ is chosen such that  
the integration over the delta function 
$\delta(\omega-k_+ -k_-)$ should yield a nonzero value. That means $\Lambda$ can be any value larger than $\omega$, but from physics
$\Lambda$ is slightly larger than $\omega$, but of the same order, i.e., $\Lambda \sim Q(1-z)$.
 Fig.~\ref{phaseDY} shows the phase space for the real gluon emission, where the shaded green region denotes 
the integration region for the part proportional to $\delta(\w)$ in Eq.~(\ref{MRDY}) or (\ref{epLdist}). 
The dashed line  represents the constraint  for  nonzero $\w$. 

Combining Eqs.~(\ref{MsV}) and (\ref{MRDY}), we obtain the naive DY soft function  at order $\alpha_s$ as 
\bea 
\tilde{\mc{S}}_{DY}^{(1)}(\omega,\mu) &=& 
\frac{\alpha_s C_F}{\pi}  \Biggl\{\delta (\omega)\Bigl[\frac{1}{\eps^2} +\frac{1}{\eps} \ln \frac{\mu^2}{\nu^2} 
 -\frac{2}{\eta} \Bigl( \frac{1}{\eps} +\ln \frac{\mu^2}{m_g^2}  \Bigr)  
 -\ln\frac{\nu^2}{\Lambda^2} \ln \frac{\mu^2}{m_g^2}  \nnb \\
\label{nDYnlo}
 &&~~~~~~+\frac{1}{2}\ln \frac{\mu^2}{\Lambda^2} -\frac{\pi^2}{4}\Bigr]+\Bigl[\frac{2}{\omega} 
 \ln \frac{\omega^2}{m_g^2}\Bigr]_{\Lambda}\Biggr\}_.
\eea  
We can clearly see that the naive one-loop result contains the IR divergence as the logarithm of $m_g$. 
This is due to the incomplete cancellation of the virtual and real corrections. If the phase space for the real gluon emission spanned all the region with
no constraint, the virtual and the real corrections would cancel, $\mc{M}_{S}^R +\mc{M}_{S}^V=0$. 
In this case, the soft function would become zero at order $\alpha_s$, 
which holds true to all orders due to the fact that $Y_n^{\dagger} Y_n = Y_{\bar{n}}^{\dagger} Y_{\bar{n}}=1$.  
However, as can be seen in Fig.~\ref{phaseDY},
the phase space for the real gluon emission does not cover all the IR region (near the red line) available to the virtual corrections near threshold. Therefore the incomplete cancellation yields IR divergence in the naive DY soft function. 
The existence of the IR divergence, which was pointed out in Ref.~\cite{Chay:2012jr,Chay:2013zya}, could invalidate the factorization near threshold. 
Furthermore, as shown in Fig.~\ref{phaseDY}, there is no rapidity divergence in the real gluon emission since the phase space responsible for the 
rapidity divergence is not included in the phase space for the real gluon emission. The rapidity divergence in the naive soft function comes solely from the
virtual contribution.
 
These problems posed by the naive soft contribution can be resolved by introducing the csoft modes. The csoft contribution is included in
the definition of the PDF, but the csoft momentum is also a subset of the soft momentum. In order to avoid double counting, the csoft contribution
is subtracted from the naive soft function to define the true soft function near threshold, given by Eq.~(\ref{nlosdys}).  
The subtraction removes both the IR and the rapidity divergences in the soft function.

Let us first consider the contribution of the $n$-csoft mode at order $\alpha_s$. In DY process, the contribution of the $\n$-csoft mode 
is the same as the $n$-csoft case due to the symmetry under $n\leftrightarrow \n$. 
We calculate the dimensionful csoft function $\mc{S}_{cs} (\w) = S_{cs}/Q$ at  order $\alpha_s$. 
Here $\w = Q(1-z)$ and the dimensionless csoft function $S_{cs}$ is defined in Eq.~(\ref{Sncs}). The virtual contribution of the csoft mode 
is the same as that of the soft mode $\mc{M}^V_{cs} = \mc{M}^V_{S}$, which is presented in Eq.~(\ref{MsV}). 

\begin{figure}[b] 
\begin{center}
\includegraphics[width=16cm]{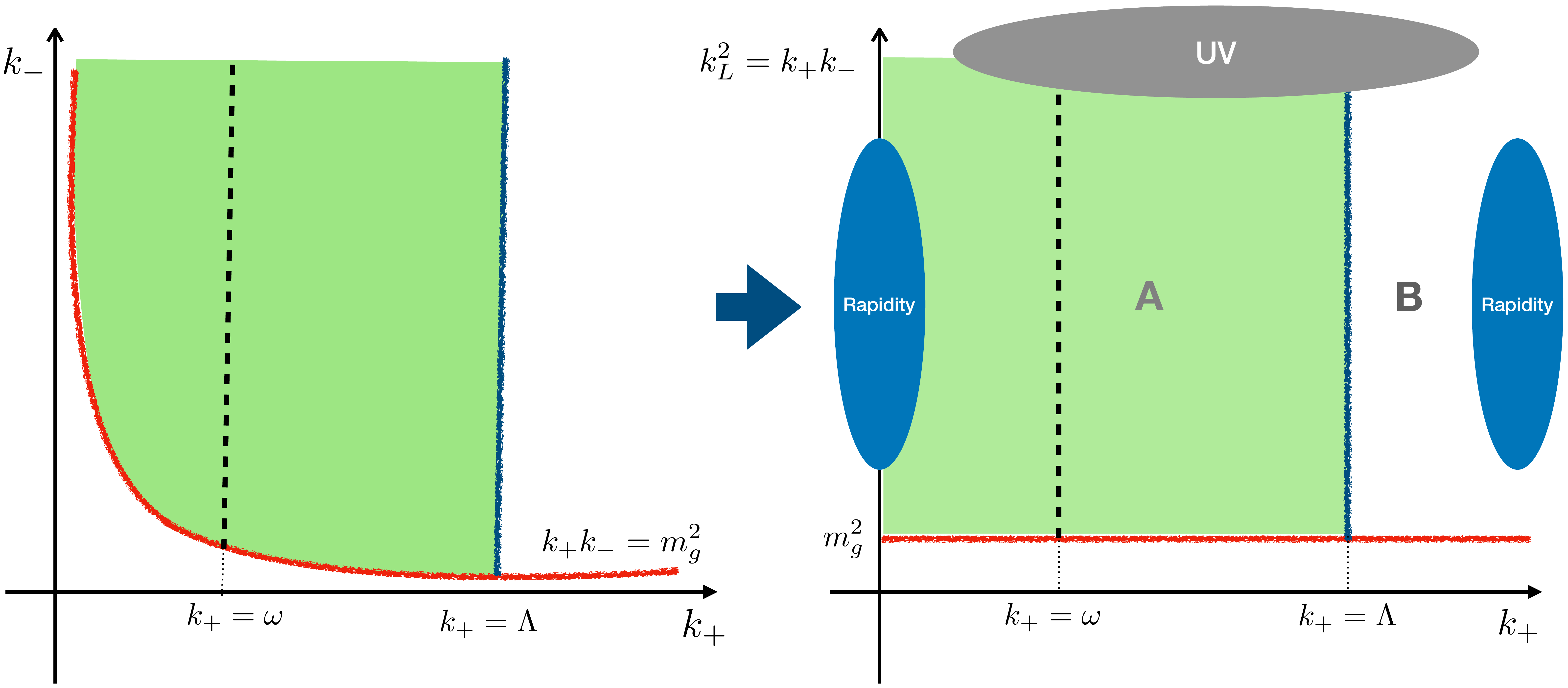}
\end{center}  
\vspace{-0.5cm}
\caption{\baselineskip 3.0 ex 
The structure of the phase space for the real $n$-csoft gluon emission. % in Eq.~(\ref{csoftR}). 
The green region is the integration region for the part proportional to $\delta(\w)$ with the $\Lambda$-distribution. 
The dotted line $k_+=\w$ is the constraint from the delta function for nonzero $\w$.  
In the $k_+$-$k_L^2$ plane, the integration region looks simple, and the regions where UV and the rapidity divergences arise are shown.
\label{fig2}}
\end{figure}

The real gluon emission at  order $\alpha_s$ is written as 
\bea
\mc{M}_{cs}^R (\w) &=& \frac{\alpha_s C_F}{\pi} \frac{(\mu^2 e^{\gamma_E})^{\eps}}{\Gamma(1-\eps)} \nu^{\eta} 
\int \frac{dk_+dk_-}{k_+k_-}(k_+k_--m_g^2)^{-\eps}|k_+ - k_-|^{-\eta} \delta (\w -k_+ ) \Theta(k_+k_--m_g^2) \nonumber \\
\label{csoftR}
&=&  \delta (\omega) \int_0^{\Lambda} d\omega^{\prime} \mc{M}_{cs}^R (\omega') 
+\Bigl[\mc{M}_{cs}^R (\omega \neq 0)\Bigr]_{\Lambda},
\eea 
where the result is written by invoking the $\Lambda$-distribution,
in which the IR divergence as $\w \to 0$ is extracted in the term proportional to $\delta(\w)$. 
In Fig.~\ref{fig2} we show the structure of the phase space for the real emission both in the $k_+$-$k_-$ and in the $k_+$-$k_L^2$ planes, 
where $k_L^2 \equiv k_+k_-$. The shaded green region is the integration region for the part proportional to $\delta(\w)$ and the dashed line is 
the phase space for $\mc{M}_{cs}^R (\omega \neq 0)$. 

When the real and the virtual contributions are combined, the  contribution of the part proportional to $\delta(\w)$ 
comes from the region B in the $k_+$-$k_L^2$ plane of Fig.~\ref{fig2} since the virtual contribution covers whole region above the line 
$k_+k_-=m_g^2$ with the minus sign relative to the real contribution. Therefore the csoft contribution at order $\alpha_s$ can be written as  
\be
\mc{S}_{cs}^{(1)}(\w,\mu) = \mc{M}_{S}^V +\mc{M}_{cs}^R = - I_B \cdot\delta(\w)+\Bigl[\mc{M}_{cs}^R (\omega \neq 0)\Bigr]_{\Lambda}.
\ee
Here $I_B$ is the integral over the region B in Fig.~\ref{fig2}, and it is given as
\bea 
I_B &=& \frac{\alpha_s C_F}{\pi} \frac{(\mu^2 e^{\gamma_E})^{\eps}}{\Gamma(1-\eps)} \nu^{\eta} \int^{\infty}_{m_g^2} dk_L^2 
\frac{(k_L^2 -m_g^2)^{-\eps}}{k_L^2}\int^{\infty}_{\Lambda} dk_+ k_+^{-1-\eta} \nnb \\
\label{calIB}
&=& \frac{\alpha_s C_F}{\pi} \left(\frac{1}{\eta}+\ln\frac{\nu}{\Lambda} \right)\left(\frac{1}{\eps}+\ln\frac{\mu^2}{m_g^2} \right)_. 
\eea
Here the original rapidity regulator $|k_+-k_-|^{-\eta}$ is replaced with $k_+^{-\eta}$. 
In the region B of Fig.~\ref{fig2}, $k_-$ in the regulator can be safely ignored since the rapidity divergence  occurs only when $k_+$ goes to infinity and $k_-$ goes to zero while $k_L^2=k_+k_-$ remains finite.
Actually keeping $k_-$ in the regulator has no effect on the calculation of the region B in the limit $\eta \to 0$. Therefore there is no difference between the original regulator and $k_+^{-\eta}$ as far as we integrate over the region B.
In the $k_+$-$k_L^2$ plane,  $I_B$ includes the UV (IR) divergence as $k_L^2$ goes to infinity ($m_g^2$). The computation of
$\mc{M}_{cs}^R (\omega \neq 0)$ is straightforward. Finally the bare csoft contribution at order $\alpha_s$ is given as 
\be
\label{csoftnlo} 
\mc{S}_{cs}^{(1)}(\w,\mu) = \frac{\alpha_s C_F}{\pi}\Biggl\{-\left(\frac{1}{\eta}+\ln\frac{\nu}{\Lambda} \right)\left(\frac{1}{\eps}+\ln\frac{\mu^2}{m_g^2} \right) \delta(\w) +\left(\frac{1}{\eps}+\ln\frac{\mu^2}{m_g^2}\right) \Bigl[\frac{1}{\w}\Bigr]_\Lambda\Biggr\}_.
\ee

The $\n$-csoft contribution $\mc{S}_{\overline{cs}}$ is the same as $\mc{S}_{cs}$ due to the symmetry under $n\leftrightarrow  \overline{n}$.
We finally obtain  the proper soft function for DY process at order $\alpha_s$ as 
\begin{eqnarray} 
\mc{S}_{\mathrm{DY}}^{(1)} (\w,\mu) &=&  \tilde{\mc{S}}_{\mathrm{DY}}^{(1)} -\mc{S}_{cs}^{(1)} - \mc{S}_{\ov{cs}}^{(1)}
= \tilde{\mc{S}}_{\mathrm{DY}}^{(1)} -2\mc{S}_{cs}^{(1)} \nonumber \\
\label{dysoft}
&=& \frac{\alpha_s C_F}{\pi} \Biggl\{\delta (\omega) \Bigl(\frac{1}{\eps^2} +\frac{1}{\eps}\ln \frac{\mu^2}{\Lambda^2} 
+\frac{1}{2} 
\ln^2 \frac{\mu^2}{\Lambda^2} -\frac{\pi^2}{4}\Bigr) 
-2\Bigl[\Bigl( \frac{1}{\eps} +\ln \frac{\mu^2}{\omega^2}\Bigr) \frac{1}{\omega}\Bigr]_{\Lambda}\Biggr\}_.
\end{eqnarray}
Note that this soft function contains only the UV divergence with neither rapidity nor IR divergences. 
This function is governed by a single scale $\w\sim \Lambda \sim Q(1-z)$, hence the scale $\mu$ to minimize 
the large logarithms is given by $Q(1-z)$. 

The dimensionful soft function in Eq.~(\ref{dysoft}) can be easily converted to the dimensionless soft function. 
From the definition of the $\Lambda$-distribution in Eq.~(\ref{Ladist}), we have the following relations to the standard plus distribution:
\begin{eqnarray} 
&&\delta (\omega) = \frac{1}{Q} \delta (1-z), \nonumber \\
\label{omiden}
&& \Bigl[\frac{1}{\omega}\Bigr]_{\Lambda} = \frac{1}{Q} \Bigl[ \frac{1}{(1-z)_+}  -\ln \frac{\Lambda}{Q} \delta(1-z)] \Bigr], 
 \\
&& \Bigl[\frac{1}{\omega} \ln \frac{\mu}{\omega}\Bigr]_{\Lambda} = \frac{1}{Q} \Bigl[\delta (1-z) \Bigl(\frac{1}{2} \ln^2 \frac{\Lambda}{Q}
-\ln \frac{\Lambda}{Q}\ln \frac{\mu}{Q}\Bigr) +\frac{1}{(1-z)_+} \ln \frac{\mu}{Q}
-\Bigl(\frac{\ln (1-z)}{1-z}\Bigr)_+ \Bigr]_. \nnb
\end{eqnarray}
Then the renormalized dimensionless soft function to next-to-leading order (NLO) in $\as$ is given as
\begin{eqnarray}
\label{dldysnlo}
S_{\mathrm{DY}} (Q(1-z),\mu) &=& \frac{1}{Q} \mc{S}_{\mathrm{DY}}(\w,\mu)  \\
= \delta(1-z)&+&\frac{\alpha_s C_F}{\pi} \Biggl\{ \delta (1-z) \Bigl[ \frac{1}{2} \ln^2 \frac{\mu^2}{Q^2} -\frac{\pi^2}{4}\Bigr] 
-2\ln \frac{\mu^2}{Q^2} \frac{1}{(1-z)_+} +4\Bigl(\frac{\ln (1-z)}{1-z}\Bigr)_+\Biggr\}_, \nnb
\end{eqnarray}
which is the same as $W_{\mathrm{DY}}$ in Ref.~\cite{Chay:2012jr,Chay:2013zya}.

We now consider the soft function in DIS. The naive soft function has been defined in Eq.~(\ref{dissoft}). Note that the naive soft function is the same as the $n$-csoft function in Eq.~(\ref{Sncs})  
except that the Wilson lines involved are the soft fields for the naive soft function, and the csoft fields for the csoft function.
The soft and $n$-csoft momenta scale as $p_s^{\mu} \sim Q\zeta(1,1,1)$ and $p_{cs}^{\mu} \sim Q\zeta (1,\alpha^2,\alpha)$. 
However, since both functions involve the same scale  $Q(1-z)\sim Q\zeta$ in the delta functions, they are the same to all orders in 
$\alpha_s$.\footnote{\baselineskip 3.0ex
In Ref.~\cite{Hoang:2015iva}, it has been shown that  the naive soft function and the csoft contribution in DIS at one loop are the same.
However, different rapidity regulators are used in the naive soft and the csoft functions in Ref.~\cite{Hoang:2015iva}. And the virtual and real contributions for each function are given  differently, but the sum turns out to be the same as ours. 
In our analysis we note that the rapidity divergence is regulated only by $k_+^{-\eta}$ 
when we combine the real and virtual contributions.
It is clear by considering the phase space analysis illustrated in Fig.~\ref{fig2}, which also holds for the naive soft function in DIS.    
}

Unlike DY process, there is no $\overline{n}$-csoft contribution in DIS. When we consider $\n$-csoft contribution from Eq.~(\ref{dissoft}),
the delta function becomes $\delta(1-z)$ since $Q(1-z)$ is  much larger than $\n\cdot p_{\ov{cs}}$ in power counting. 
The $\n$-csoft real emission is the same as the virtual
$\overline{n}$-csoft contribution but with the opposite sign. Therefore the $\overline{n}$-csoft contribution cancels at one loop, and to all orders 
in $\alpha_s$. As a result, the proper soft function in DIS remains as $\delta(1-z)$ to all orders in $\as$.

\subsection{The PDF at NLO near threshold}

From the definition of the PDF near threshold in Eq.~(\ref{pdflx}), we compute the correction at order $\alpha_s$.
 If we take the perturbative limit, the PDF at the parton level can be additionally factorized 
as~\cite{Becher:2006mr,Fleming:2012kb,Hoang:2015iva,Korchemsky:1992xv}
\bea 
\phi_{q/q}(x,\mu) &=& \frac{1}{2Q} \mr{Tr} \langle q (p_+) | \bar{\chi}_n | 0 \rangle\nn  \langle 0 | \chi_n |q(p_+)\rangle \cdot 
\frac{1}{N_c} \mathrm{Tr}  \langle 0|Y_{n,cs}^{\dagger} Y_{\bar{n},cs} 
\delta \Bigl( 1-x+\frac{\n\cdot i\partial}{Q} \Bigr) Y_{\bar{n},cs}^{\dagger} Y_{n,cs}|0\rangle \nnb \\
\label{facpdf}
&=& \mc{C}_q (Q,\mu) \cdot S_{cs} (Q(1-x),\mu), 
\eea
where $Q=p_+$ is the large quark momentum, $\mc{C}_q$ is the collinear part, and $S_{sc}$ is the csoft function defined in Eq.~(\ref{Sncs}).   

At one loop, the virtual contribution for $\mc{C}_q$ is given as 
\be
\label{Mcloop}
\mc{M}_C^V = \frac{\alpha_s C_F}{\pi} \Bigl[\Bigl(\frac{1}{\eps} +\ln \frac{\mu^2}{m_g^2}\Bigr)\Bigl( \frac{1}{\eta}+
\ln \frac{\nu}{Q} +1\Bigr) +1-\frac{\pi^2}{6}\Bigr]_.
\ee 
The wave function renormalization and the residue are given by 
\be
\label{self}
Z_q^{(1)} + R_q^{(1)} = -\frac{\as}{4\pi}\Bigl(\frac{1}{\eps}+\ln \frac{\mu^2}{m_g^2}-\frac{1}{2}\Bigr)_.
\ee
And we obtain  the collinear part to NLO as 
\bea
\label{Cqnlo}
\mc{C}_q (Q,\mu) &=& 1+\mc{M}_C^V +Z_q^{(1)} + R_q^{(1)} \nnb \\
&=& 1+ \frac{\alpha_s C_F}{\pi} \Bigl[\Bigl(\frac{1}{\eps} +\ln \frac{\mu^2}{m_g^2}\Bigr)\Bigl( \frac{1}{\eta}+
\ln \frac{\nu}{Q} +\frac{3}{4}\Bigr) +\frac{9}{8}-\frac{\pi^2}{6}\Bigr]_.
\eea 
It contains the rapidity,  the UV and the IR divergences.

From Eqs.~(\ref{csoftnlo}) and (\ref{omiden}), we also obtain the dimensionless csoft function to NLO as 
\bea
\label{dlScsnlo}
S_{cs} (Q(1-x),\mu) &=& \delta(1-x)+ \frac{\as C_F}{\pi} \Biggl\{-\Bigl(\frac{1}{\eps} +\ln \frac{\mu^2}{m_g^2}\Bigr)
\Bigl( \frac{1}{\eta}+\ln \frac{\nu}{Q} \Bigr)\delta(1-x) \nnb \\
&&~~~~~~~~~~~~+\Bigl(\frac{1}{\eps} +\ln \frac{\mu^2}{m_g^2}\Bigr)\frac{1}{(1-x)_+}\Biggr\}_. 
\eea
Therefore the bare PDF to order $\alpha_s$ is given as 
\be \label{PDFnlo} 
\phi_{q/q}(x,\mu) = 1+\frac{\as C_F}{2\pi}\Biggl\{\delta(1-x) \Biggl[\frac{3}{2}\Bigl(\frac{1}{\eps} 
+\ln \frac{\mu^2}{m_g^2}\Bigr)+\frac{9}{4}-\frac{\pi^2}{3} \Biggr]  
 + \frac{2}{(1-x)_+}\Bigl(\frac{1}{\eps} +\ln \frac{\mu^2}{m_g^2}\Bigr)\Biggr\}_.
\ee
The PDF near threshold is free of rapidity divergence, and it is the same as the PDF away from threshold when we take the limit $x\to 1$. 
Obviously, the renormalization-group behavior satisfies the DGLAP evolution in the limit $x\to 1$, and we do not repeat solving the renormalization
group equations and refer to Refs.~\cite{Becher:2006mr,Becher:2007ty}.

\section{Dihadron production in $e^+ e^-$ annihilation near threshold }
\label{dihadron} 

We consider the dihadron production in $e^+ e^-$ annihilation near threshold: $e^-   e^+ \rightarrow h_1  +h_2  +X$, where $X$ denotes soft 
particles in the final state. Here the final hadrons $h_1$ and $h_2$ in the $n$ and $\overline{n}$ direction take almost all the 
energies of the mother partons $p_{h1}^+ = x_1 Q$,
 $p_{h2}^- = x_2 Q$, where $Q$ is the center-of-mass energy and $x_{1,2}$ are close to 1. 
The scattering cross section is factorized into the two fragmentation functions and the soft function.   
If we naively compute the soft function without the csoft subtraction, here we also have the IR divergence as well as the rapidity divergence,
invalidating the factorization theorem. Therefore we define the fragmentation functions in terms of the collinear and the csoft fields 
describing the radiations in the directions of the observed hadrons. Then we can properly subtract the csoft interactions from 
the naive soft function, and as a result the factorization theorem can be written as
\begin{equation}
\label{facDH}
\frac{d\sigma}{dp_{h1}^+ dp_{h2}^-} = \frac{\sigma_0 H_{\mr{DH}}(Q^2,\mu)}{Q^2} \int_{x_1}^1 \frac{dz_1}{z_1} \int_{x_2}^1 \frac{dz_2}{z_2} D_{h_1/q} \Bigl(\frac{x_1}{z_1},\mu\Bigr)
D_{h_2/\bar{q}} \Bigl(\frac{x_2}{z_2},\mu\Bigr) S_{\mr{DH}} (Q(1-z_1), Q(1-z_2),\mu).
\end{equation}
Here $\sigma_0$ is the Born scattering cross section and the threshold region corresponds to $x_{1,2} \to 1$. 
The hard function $H_{\mr{DH}}$ is given by the same as $H_{\mr{DY}}$ in Eq.~(\ref{fdyc}). 
$S_{\mr{DH}}$ is the soft function for the dihadron production and it can be obtained after the csoft subtraction from the naive soft function. 

Following the analysis of the fragmentation function to a jet (FFJ) in the large $z$ limit~\cite{Dai:2017dpc}, we  define the hadron fragmentation 
function as
\begin{equation} 
\label{fragth}
D_{h_1/q} (z_1,\mu) = \frac{z_1^{D-3}}{2N_c} \mathrm{Tr} \langle 0| 
Y_{\bar{n},cs}^{\dagger} Y_{n,cs} \frac{\overline{\FMslash{n}}}{2} \chi_n |h_1\rangle \langle h_1| \overline{\chi}_n 
\delta \Bigl( \frac{p_{h_1}^+}{z_1} -\mathcal{P}_+^{\dagger} +\overline{n}\cdot i\partial
\Bigr)Y_{n,cs}^{\dagger}Y_{\bar{n},cs} |0\rangle,
\end{equation}
where we set the transverse momentum of the hadron $\blpu{p}_{h1}$ as zero.
Similarly, the fragmentation function from the antiquark $D_{h_2/\bar{q}} (z_2,\mu)$ can be obtained in terms of the $\n$-collinear and 
$\n$-csoft fields. As in the case of the PDF, this definition is valid near, and away from threshold.
Near threshold, putting $\mc{P}_+^{\dagger} = p_{h_1}^+$ we can further simplify Eq.~(\ref{fragth}) as  
\begin{equation} 
\label{fragthz}
D_{h_1/q} (z_1\to 1,\mu) = \frac{1}{2N_c} \mathrm{Tr} \langle 0| 
Y_{\bar{n}}^{cs\dagger} Y_n^{cs} \frac{\overline{\FMslash{n}}}{2} \chi_n |h_1\rangle \langle h_1| \overline{\chi}_n 
\delta ( p_{h_1}^+(1-z_1) +\overline{n}\cdot i\partial )Y_{n,cs}^{\dagger}Y_{\bar{n},cs} |0\rangle.
\end{equation}
We can easily check that the result at order $\alpha_s$ for Eq.~(\ref{fragthz}) at the parton level is given by 
the same as the result on the PDF and its renormalization behavior follows the DGLAP evolution in the large $z$ limit. 

For the soft function, we start with the naive soft function, which is defined as 
\be
\label{nsoftdh} 
\tilde{S}_{\mr{DH}}\bigl(Q(1-z_1), Q(1-z_2)\bigr) = \frac{1}{N_c} \mathrm{Tr} \langle 0| Y_{\bar{n}}^{\dagger} Y_n 
\delta \Bigl(1-z_1 +\frac{\n\cdot i\partial}{Q}\Bigr) \delta \Bigl( 1-z_2 +\frac{n\cdot i\partial}{Q}\Bigr) Y_n^{\dagger} Y_{\bar{n}}|0\rangle.
\ee
The csoft functions can be obtained by taking the $n$- and $\n$-csoft limits on $\tilde{S}_{\mr{DH}}$, which are subtracted from the naive soft function. 
As a result the NLO correction to the soft function for dihadron production is given by
\be
\label{nlosdhs} 
S_{\mathrm{DH}}^{(1)}  = \tilde{S}_{\mathrm{DH}}^{(1)} - S_{cs}^{(1)}\cdot\delta(1-z_2) - S_{\overline{cs}}^{(1)} \cdot 
\delta(1-z_1),
\ee
where the csoft functions $S_{cs}$ and  $S_{\overline{cs}}$ have been defined in Eqs.~(\ref{Sncs}) and (\ref{Snbcs}) respectively.
 
In order to see the scale dependence transparently, we consider the NLO calculation with the dimensionful soft function. 
It is defined as $\mc{S}_{\mr{DH}} (\w_+,\w_-) = S_{\mr{DH}}(Q(1-z_1), Q(1-z_2))/Q^2$, where $\w_+ = Q(1-z_1)$ and $\w_- = Q(1-z_2)$. 
Then the naive dimensionful soft function is expressed as 
\begin{equation}
\label{nsoftdhdef}
\tilde{\mc{S}}_{\mr{DH}}(\w_+, \w_-) =\frac{1}{N_c} \mathrm{Tr} \langle 0|  Y_{\bar{n}}^{\dagger} Y_n 
\delta (\w_+ +\overline{n}\cdot i\partial)
\delta (\w_- +n\cdot i\partial)  Y_n^{\dagger} Y_{\bar{n}}|0\rangle.
\end{equation}

The virtual one-loop contribution to $\tilde{\mc{S}}_{\mr{DH}}$ is given by $\mc{M}_{S,\mr{DH}}^V=\delta(\w_+)\delta(\w_-)\bar{\mc{M}}_{S}^V $, where   the one loop result $\bar{\mc{M}}_{S}^V$ is given in Eq.~(\ref{MsV}). 
And the real contribution is given as
\begin{eqnarray}
\mc{M}_{S,\mr{DH}}^R &=& \frac{\alpha_s C_F}{\pi} \frac{(\mu^2 e^{\gamma_E})^{\eps}}{\Gamma(1-\eps)} \nu^{\eta} \int \frac{dk_+dk_-}{k_+k_-}(k_+k_--m_g^2)^{-\eps}|k_+ - k_-|^{-\eta} \Theta(k_+k_--m_g^2)\nonumber \\
&&~~~~~~~~~\times \delta (\w_+ -k_+ )\delta (\w_- -k_- ) \nnb \\
\label{MRDH}
&=& \frac{\alpha_s C_F}{\pi}  \frac{\Theta(\w_+\w_--m_g^2)}{\w_+\w_-}_. 
\end{eqnarray}
Here we put $\eps=\eta =0$  since the integral has no UV, rapidity divergences. 

$\mc{M}_{S,\mr{DH}}^R$ in Eq.~(\ref{MRDH}) is IR divergent as $\w_{\pm} \to 0$. In order to extract the IR divergence, 
we employ the $\Lambda$-distribution defined in Eq.~(\ref{Ladist}), where the upper values for $\w_{\pm}$ are set as $\Lambda_{\pm}$. 
Then Eq.~(\ref{MRDH}) can be rewritten as 
\begin{eqnarray}
\mc{M}_{S,\mr{DH}}^R &=& \frac{\alpha_s C_F}{\pi}\Biggl\{\delta (\w_+) \delta (\w_-) \int_{m_g^2/\Lambda_-}^{\Lambda_+} \frac{dk_+}{k_+} \int_{m_g^2/k_+}^{\Lambda_-}\frac{dk_-}{k_-} 
+\delta (\w_-)\Bigl[\frac{1}{\w_+} \int_{m_g^2/\w_+}^{\Lambda_-} \frac{dk_-}{k_-}\Bigr]_{\Lambda_+}   \nonumber \\
&&~~~~~+ \delta (\w_+) \Bigl[\frac{1}{\w_-} \int_{m_g^2/\w_-}^{\Lambda_+} \frac{dk_+}{k_+}\Bigr]_{\Lambda_-} 
+\Bigl[\frac{1}{\w_+}\Bigr]_{\Lambda_+} \Bigl[\frac{1}{\w_-}\Bigr]_{\Lambda_-}\Biggr\} \nonumber \\
&=&  \frac{\alpha_s C_F}{\pi}\Biggl\{\delta (\w_+) \delta (\w_-)  \frac{1}{2} \ln^2 \frac{\Lambda_+ \Lambda_-}{m_g^2} + \delta (\w_-)\Bigl[\frac{1}{\w_+} 
\ln \frac{\Lambda_- \w_+}{m_g^2}\Bigr]_{\Lambda_+}  \nonumber \\
&&~~~~~+\delta (\w_+)\Bigl[\frac{1}{\w_-} 
\ln \frac{\Lambda_+ \w_-}{m_g^2}\Bigr]_{\Lambda_-}  +\Bigl[\frac{1}{\w_+}\Bigr]_{\Lambda_+}
\Bigl[\frac{1}{\w_-}\Bigr]_{\Lambda_-}\Biggr\}_. 
\end{eqnarray}

Combining the virtual and the real contributions, we obtain the naive soft function at order $\alpha_s$ as 
\bea 
\tilde{\mc{S}}_{\mr{DH}}^{(1)}(\w_+, \w_-) &=&  \delta(\w_+)\delta(\w_-) \bar{\mc{M}}_{S}^V + \mc{M}_{S,\mr{DH}}^R (\w_+,\w_-) \nnb \\
&=& \frac{\alpha_s C_F}{\pi}  \Biggl\{\delta (\omega)\Bigl[\frac{1}{\eps^2} +\frac{1}{\eps} \ln \frac{\mu^2}{\nu^2} 
 -\frac{2}{\eta} \Bigl( \frac{1}{\eps} +\ln \frac{\mu^2}{m_g^2} -\ln\frac{\nu^2}{m_g^2} \ln \frac{\mu^2}{m_g^2} \Bigr)  
 +\frac{1}{2}  \ln^2 \frac{\mu^2}{m_g^2}  \nnb \\
&&~~~~~+\frac{1}{2}  \ln^2 \frac{\Lambda_+\Lambda_-}{m_g^2} -\frac{\pi^2}{12}\Bigr]+ \delta (\w_-)\Bigl[\frac{1}{\w_+} 
\ln \frac{\Lambda_- \w_+}{m_g^2}\Bigr]_{\Lambda_+} \nnb \\
\label{nlonsdh}
&&~~~~~+\delta (\w_+)\Bigl[\frac{1}{\w_-} 
\ln \frac{\Lambda_+ \w_-}{m_g^2}\Bigr]_{\Lambda_-}  +\Bigl[\frac{1}{\w_+}\Bigr]_{\Lambda_+}
\Bigl[\frac{1}{\w_-}\Bigr]_{\Lambda_-}\Biggr\}_. 
\eea
This result contains the IR and the rapidity divergences as  expected. 

From Eq.~(\ref{nlosdhs}) the csoft contributions to be subtracted from the naive dimensionful soft function are given by 
$\delta(\w_-)\mc{S}_{cs}^{(1)}(\w_+)+\delta(\w_+)\mc{S}_{\ov{cs}}^{(1)}(\w_-)$. And using Eq.~(\ref{csoftnlo}) we write the NLO csoft 
contributions as 
\bea
\delta(\w_-)\mc{S}_{cs}^{(1)}(\w_+)+\delta(\w_+)\mc{S}_{\ov{cs}}^{(1)}(\w_-) &=& 
\frac{\alpha_s C_F}{\pi}\Biggl\{-\left(\frac{2}{\eta}+\ln\frac{\nu^2}{\Lambda_+\Lambda_-} \right)\left(\frac{1}{\eps}+\ln\frac{\mu^2}{m_g^2} \right) \delta(\w_+) \delta(\w_-) \nnb \\
\label{dhcsoft}
&&\hspace{-1cm}
+\left(\frac{1}{\eps}+\ln\frac{\mu^2}{m_g^2}\right) \left(\delta(\w_-)\Bigl[\frac{1}{\w_+}\Bigr]_{\Lambda_+}
+\delta(\w_+)\Bigl[\frac{1}{\w_-}\Bigr]_{\Lambda_-}\right)\Biggr\}_.
\eea

Finally, subtracting Eq.~(\ref{dhcsoft}) from Eq.~(\ref{nlonsdh}) we obtain the NLO result for the bare dimensionful soft function as 
\bea
\mc{S}_{\mr{DH}}(\w_+,\w_-,\mu) &=& \delta(\w_+)\delta(\w_-) + \frac{\alpha_s C_F}{\pi} \Biggl\{ \delta(\w_+)\delta(\w_-)
\Bigl[\frac{1}{\eps^2} +\frac{1}{\eps} \ln \frac{\mu^2}{\Lambda_+ \Lambda_-}
+\frac{1}{2} \ln^2 \frac{\mu^2}{\Lambda_+ \Lambda_-} -\frac{\pi^2}{12}\Bigl] \nnb \\
&&~~~~~
-\delta (\w_-) \Bigl[\frac{1}{\w_+} \Bigl(\frac{1}{\eps} +\ln \frac{\mu^2}{\Lambda_-\w_+}\Bigr) \Bigr]_{\Lambda_+}
-\delta (\w_+) \Bigl[\frac{1}{\w_-} \Bigl(\frac{1}{\eps} +\ln \frac{\mu^2}{\Lambda_+\w_-}\Bigr) \Bigr]_{\Lambda_-}\nnb\\
&&~~~~~~
\label{nlosdh}
+\Bigl[\frac{1}{\w_+}\Bigr]_{\Lambda_+}\Bigl[\frac{1}{\w_-}\Bigr]_{\Lambda_-}\Biggr\}_.
\eea
We clearly see that the problematic IR and rapidity divergences are removed by the csoft subtraction as in DY process. 
Using Eq.~(\ref{omiden}), we can convert Eq.~(\ref{nlosdh}) to the dimensionless soft function. And the scales for the logarithms are determined
by $\mu \sim \Lambda_{\pm}$. As a result, the renormalized soft function at NLO is given as 
\bea
S_{\mr{DH}}\bigl(Q(1-z_1),Q(1-z_2),\mu\bigr) &=& \delta(1-z_1)\delta(1-z_2) + \frac{\alpha_s C_F}{\pi} \Biggl\{ \delta(1-z_1)\delta(1-z_2)\Bigl(\frac{1}{2}\ln\frac{\mu^2}{Q^2}-\frac{\pi^2}{12}\Bigr) \nnb\\
&&\hspace{-3cm} 
-\delta(1-z_2)\Bigl[\frac{1}{1-z_1}\ln\frac{\mu}{Q(1-z_1)}\Bigr]_+ 
-\delta(1-z_1)\Bigl[\frac{1}{1-z_2}\ln\frac{\mu}{Q(1-z_2)}\Bigr]_+ \nnb \\
&&\hspace{-3cm} 
\label{nlodmsdh}
+\Bigl[\frac{1}{1-z_1}\Bigr]_+ \Bigl[\frac{1}{1-z_2}\Bigr]_+. 
\eea

\section{Conclusion\label{conc}}

In the framework of SCET, we have scrutinized the factorization theorems near threshold in DY, DIS processes and in the dihadron production in $e^+ e^-$ annihilation 
by introducing the csoft modes in SCET. The important point in analyzing these processes near threshold is that there appears a csoft mode governed by the scale $\omega =Q(1-z) \ll Q$. 
Near threshold, real collinear particles for the PDF cannot be emitted due to the kinematical constraint, 
while the csoft modes can. 
The effect of the csoft modes can be implemented by decoupling the csoft modes from
the collinear fields. The resultant PDF consists of the csoft Wilson lines along with the collinear fields.  
Note that the definitions of the PDF and the fragmentation function in
Eqs.~(\ref{pdfth}) and (\ref{fragth}) with the csoft modes are valid not only near threshold but also away from threshold. Away from threshold, the effect of the csoft
mode is cancelled, and the collinear mode alone describes the whole process. 

The naive collinear contribution to the PDF contains the rapidity divergence, but it is cancelled when the contribution of 
the csoft modes is included. And the resultant PDFs are the PDF obtained from the full QCD. 
In Ref.~\cite{Chay:2013zya}, the same conclusion has been obtained by reshuffling suitable divergences in the collinear and the soft parts, 
but here the physics becomes more elaborate. 
It is also true for the fragmentation function that the definition including the csoft modes yields the fragmentation function in the full QCD, which
can be extended to the threshold region.

The csoft mode is also a subset of the soft mode, hence the contribution of the csoft modes  should be subtracted in the soft part to avoid
double counting. And the soft functions after the subtraction are free of IR and rapidity divergences, and can be handled perturbatively. 
It is because the phase space for the soft modes responsible
for the IR and rapidity divergences coincides with the phase space for the csoft modes. Therefore the IR and rapidity divergences are cancelled when 
the csoft modes are subtracted from the soft part. 

The renormalization group evolution of the soft functions and the PDF can be obtained, for example, as in Refs.~\cite{Becher:2006mr,Becher:2007ty}, 
so we will not repeat it here. But the important point is that the introduction of the csoft modes justifies the use of the renormalization group 
equation since we have explicitly verified that the higher-order correction to the soft function in DY process (it is zero in DIS) is indeed IR finite and free of rapidity divergences. 
So far, it has been assumed that the soft functions should be IR finite. But, in fact, the removal of the IR and rapidity divergences is sophisticated and
it is accomplished by including the csoft modes. In this paper, we give a firm basis to the use of the renormalization group equation 
in resumming large logarithms near threshold.

In Refs.~\cite{Fleming:2012kb,Fleming:2016nhs}, the authors have considered the same processes, i.e., DY and DIS processes near threshold.
However, their results are different from the ones
presented here. The main difference is that the PDFs in Refs.~\cite{Fleming:2012kb,Fleming:2016nhs} are formulated by combining the collinear and the soft mode. As a result there arises a correlation between the two collinear sectors in DY process. The prescription in Refs.~\cite{Fleming:2012kb,Fleming:2016nhs} might hold only when $Q\zeta \sim \w$ is close to $\Lambda_{\mr{QCD}}$, where the csoft modes become identical to the soft mode.  
In this paper we set $Q\zeta \sim \w$ as a free small energy scale. As far as $Q\zeta \sim \w \gg \Lambda_{\mr{QCD}}$, with the help of the csoft modes, we can formulate the factorization theorem where no correlation arises between two collinear and soft sectors.

The final results presented here are the same as those in previous literature~\cite{Chay:2012jr,Chay:2013zya}, so we may wonder what can be learned from this analysis. 
In spite of the same result, there are some illuminating points in our factorization procedure, which are worth commenting. 
First, the new scale
$Q(1-z)$ is introduced, and it does not have to be related to $\Lambda_{\mr{QCD}}$.
Previously it has been considered
to accommodate the new scale with the power counting of $\lambda \sim \Lambda_{\mr{QCD}}/Q$.

Second, the introduction of the csoft modes yields 
the soft function free of IR and rapidity divergences. Without the csoft modes, careful analysis shows that the real emission in the soft function
contains IR divergence which is not cancelled by the virtual correction. Furthermore, there exists rapidity divergence. We have confirmed that
the soft functions are indeed free of IR and rapidity divergences, and the PDF turns out to be the same as the PDF in full QCD. 

Third, it becomes apparent which scale governs in each factorized parts. The naive soft function includes the contributions from
different scales, but by separating the csoft modes, it now becomes apparent that each part receives the appropriate  scale dependence. For the 
PDF, the scale in the logarithm is of order $\mu \sim \Lambda_{\mathrm{QCD}}$ or larger and for the soft function, 
$\mu \sim \omega = Q(1-z)$, or $Q(1-x)$. From the study of the dijet production~\cite{Becher:2015hka,Chien:2015cka}, we now know that there should be additional 
degrees of freedom to account for the resummation with respect to the different scales in different factorized parts.  
And it is also true in DY and DIS processes, as well as the dihadron production, near threshold.

In conclusion, we have analyzed the factorization near threshold in SCET by including the csoft modes in defining the PDF, and subtracting the csoft contributions for the soft functions. 
The newly defined PDF can be properly extended to threshold and the resultant factorization theorem takes care of the problem of the IR and rapidity divergences in the soft function, which enables the resummation of the large logarithms through the renormalization group equation. Only after the inclusion of the csoft modes, the factorized result in SCET, Eqs.~(\ref{fdyc}) and (\ref{fdisc}), is consistent with the 
result in full QCD.

\begin{acknowledgments}
J. Chay is supported by Basic Science Research Program through the National Research Foundation of Korea (NRF) funded by 
the Ministry of Education(Grant No. NRF-2016R1D1A1B03935799). C.~Kim was supported by Basic Science Research Program through the National Research Foundation of Korea (NRF) funded by the Ministry of Science and ICT (Grants No. NRF-2014R1A2A1A11052687, No. NRF-2017R1A2B4010511).
\end{acknowledgments}

\end{document}